\documentclass[aps,showpacs,epsfig,twocolumn]{revtex4}
\usepackage{epsfig}

\newcommand{\be}{\begin{equation}}
\newcommand{\ee}{\end{equation}}
\newcommand{\bea}{\begin{eqnarray}}
\newcommand{\eea}{\end{eqnarray}}
\usepackage{color}
\begin{document}

\title{\bf\Large {Two ferromagnetic phases in spin-Fermion systems}}

\author{Naoum Karchev }

\affiliation{Division of Material Physics, Graduate School of
Engineering Science, Osaka University, Toyonaka, Osaka 560-8531,
Japan \\
Department of Physics, University of Sofia, 1126 Sofia, Bulgaria }

\begin{abstract}
We consider spin-Fermion systems which obtain their magnetic properties
from a system of localized magnetic moments being coupled to
conducting electrons. The dynamical degrees of freedom are spin-$s$
operators of localized spins and spin-$1/2$ Fermi operators of
itinerant electrons. We develop modified spin-wave theory and obtain
that system has two ferromagnetic phases. At the characteristic
temperature $T^{*}$, the magnetization of itinerant electrons becomes
zero, and high temperature ferromagnetic phase ($T^{*}<T<T_C$) is a
phase where only localized electrons give contribution to the
magnetization of the system. An anomalous increasing of
magnetization below $T^{*}$ is obtained in good agrement with
experimental measurements of the ferromagnetic phase of $UGe_2$.
\end{abstract}

\pacs{75.30.Et, 71.27.+a, 75.10.Lp, 75.30.Ds} \maketitle

This Brief Report is inspired from the experimental measurements of the
ferromagnetic phase of $UGe_2$ which reveal the presence of an
additional phase line that lies entirely within the ferromagnetic
phase. The characteristic temperature of this transition $T^{*}$,
which is below the Curie temperature $T_C$, decreases with pressure
and disappears at a pressure close to the pressure at which new
phase of coexistence of superconductivity and ferromagnetism
emerges\cite{2fmp1,2fmp2,2fmp3}.Strong anomaly in resistivity is
observed at $T^{*}$ \cite{2fmp4}. The additional phase transition
demonstrates itself and through the change in the $T$ dependence of
the ordered ferromagnetic moment\cite{2fmp2,2fmp5,2fmp6,2fmp7}. The
magnetization shows an anomalous enhancement below $T^{*}$. An
anomaly is found in the heat capacity at the characteristic
temperature $T^{*}$ \cite{2fmp8}. Theoretically, it was assumed that the interplay of
charge-density wave and spin-density wave fluctuations is the origin of anomalous properties\cite{2fmp8a}.
Alternatively, it was proposed that the unusual phase diagram is result
of novel tuning of the Fermi surface topology by the magnetization\cite{2fmp8b}.

Our objective is spin-Fermion systems, which obtain their magnetic
properties from a system of localized magnetic moments and itinerant electrons.
It is obtained that the true magnons in these systems, which are
the transversal fluctuations corresponding
to the total magnetization, are complicate mixture of the transversal
fluctuations of the spins of localized and itinerant electrons. The magnons
interact with different spins in a
different way, and magnons' fluctuations suppress the ordered moments
of the localized and itinerant electrons at different temperatures. As a result, the
ferromagnetic phase is divided onto two phases: low temperature
phase  $0<T<T^{*}$, where all electrons contribute the ordered ferromagnetic
moment, and high temperature phase $T^{*}<T<T_C$, where only localized spins form
magnetic moment. To describe the two phases, a modified spin-wave theory is developed.
We have reproduced theoretically the anomalous temperature dependence
of the ordered moment, known from the experiments with $UGe_2$
\cite{2fmp2,2fmp5,2fmp6,2fmp7}.

The spin-fermion model is known as $s-d$ (or $s-f$). The model
appears in the literature also as the ferromagnetic Kondo Lattice
model (FKLM) or the double exchange model (DEM). The exact results
for the spin-Fermion model are reported in \cite{2fmp8c}.

The dynamical degrees of freedom are spin-$s$ operators of localized
spins and spin-$1/2$ Fermi operators of itinerant electrons.
 We consider a theory with Hamiltonian \bea \label{letter1}\nonumber
h & = & H-\mu N = -t\sum\limits_{  \langle  ij
 \rangle  } {\left( {c_{i\sigma }^ + c_{j\sigma } + h.c.} \right)}
  -\mu \sum\limits_i {n_i
} \\ & - & J^l\sum\limits_{  \langle  ij  \rangle  } {{\bf S}_i
\cdot {\bf S}_j}
  - J\sum\limits_i {{\bf
S}_i}\cdot {\bf s}_i \eea where $s^{\nu}_i=\frac 12
\sum\limits_{\sigma\sigma'}c^+_{i\sigma}
\tau^{\nu}_{\sigma\sigma'}c^{\phantom +}_{i\sigma'}$, with the Pauli
matrices $(\tau^x,\tau^y,\tau^z)$, is the spin of the conduction
electrons, ${\bf S}_i$ is the spin of the localized electrons, $\mu$
is the chemical potential, and $n_i=c^+_{i\sigma}c_{i\sigma}$. The
sums are over all sites of a three-dimensional cubic lattice, and
$\langle i,j\rangle$ denotes the sum over the nearest neighbors. The
Heisenberg term $(J^l > 0)$ describes ferromagnetic Heisenberg
exchange between nearest-neighbors localized electrons. The term
which describes the spin-Fermion interaction $(J>0)$ is known as a
Kondo interaction $(J=J_K)$ in the ferromagnetic Kondo model or as a
Hund's term in the double exchange model ($J=J_H$ and $J^l< 0$).

We represent the Fermi operators in terms of the Schwinger bosons
($\varphi_{i,\sigma}, \varphi_{i,\sigma}^+$) and slave Fermions
($h_i, h_i^+,d_i,d_i^+$). The Bose fields
%$\varphi_{i,\sigma},\varphi_{i,\sigma}^+$
are doublets $(\sigma=1,2)$ without charge, while Fermions
%$h_i,h_i^+,d_i,d_i^+$
are spinless with charges 1 ($d_i$) and -1 ($h_i$).
\begin{eqnarray} & & c_{i\uparrow} =
h_i^+\varphi _{i1}+ \varphi_{i2}^+ d_i, \qquad c_{i\downarrow} =
h_i^+ \varphi _{i2}- \varphi_{i1}^+ d_i, \nonumber
\\
& & n_i = 1 - h^+_i h_i +  d^+_i d_i,\quad  s^{\nu}_i=\frac 12
\sum\limits_{\sigma\sigma'} \varphi^+_{i\sigma}
{\tau}^{\nu}_{\sigma\sigma'} \varphi_{i\sigma'},\nonumber
\\& &
\varphi_{i1}^+ \varphi_{i1}+ \varphi_{i2}^+ \varphi_{i2}+ d_i^+
d_i+h_i^+ h_i=1 \label{letter2}
\end{eqnarray}

 Next, we make a change of variables, introducing 
Bose doublets $\zeta_{i\sigma}$ and
$\zeta^+_{i\sigma}\,$\cite{2fmp9}
\begin{eqnarray}
\zeta_{i\sigma} & = & \varphi_{i\sigma} \left(1-h^+_i h_i-d^+_i
d_i\right)^
{-\frac 12},\nonumber \\
\zeta^+_{i\sigma} & = & \varphi^+_{i\sigma} \left(1-h^+_i h_i-d^+_i
d_i\right)^ {-\frac 12}, \label{letter6}
\end{eqnarray}
where the new fields satisfy the constraint
$\zeta^+_{i\sigma}\zeta_{i\sigma}\,=\,1$. In terms of the new fields
the spin vectors of the itinerant electrons have the form \be
s^{\nu}_{i}=\frac 12 \sum\limits_{\sigma\sigma'} \zeta^+_{i\sigma}
{\tau}^{\nu}_{\sigma\sigma'} \zeta_{i\sigma'} \left[1-h^+_i
h_i-d^+_i d_i\right] \label{letter8} \ee When, in the ground state,
the lattice site is empty, the operator identity $h^+_ih_i=1$ is
true. When the lattice site is doubly occupied, $d^+_id_i=1$. Hence,
when the lattice site is empty or doubly occupied the spin on this
site is zero. When the lattice site is neither empty nor doubly
occupied ($h^+_ih_i=d^+_id_i=0$), the spin equals $\,\,{\bf s}_{i}=1/2
{\bf n}_i,\,\,$ where the unit vector $\,\,
n^{\nu}_i=\sum\limits_{\sigma\sigma'} \zeta^+_{i\sigma}
{\tau}^{\nu}_{\sigma\sigma'} \zeta_{i\sigma'}\,\,\, ({\bf
n}_i^2=1)\,\,$ identifies the local orientation of the spin of the
itinerant electron. Let us average the spin of electrons in the
subspace of the Fermions $(d^+_i, \,d_i)$ and $(h^+_i,\,h_i)$ (to
integrate the Fermions out in the path integral approach) and
introduce the notation
\begin{equation}
m=\frac 12 \left(1-<h^+_i h_i>-<d^+_id_i>\right). \label{letter9}
\end{equation}
One obtains ${\bf s}_{i}=m {\bf n}_i$ where ${\bf s}_{i}^2=m^2$.
Hence, the amplitude of the spin vector $m$ is an effective spin of
the itinerant electrons accounting for the fact that some sites, in
the ground state, are doubly occupied or empty.

It is more convenient to use the rescaled Bose fields
\begin{equation}
\xi_{i\sigma}=\sqrt{2m}\,\zeta_{i\sigma},\qquad\qquad
\xi^+_{i\sigma}=\sqrt{2m}\,\zeta^+_{i\sigma} \label{letter10}
\end{equation}
which satisfy the constraint $\xi^+_{i\sigma}\xi_{i\sigma}=2m$. Let
us introduce the vector,
\begin{equation}
 M^{\nu}_{i}=\frac 12 \sum\limits_{\sigma\sigma'} \xi^+_{i\sigma}
{\tau}^{\nu}_{\sigma\sigma'} \xi_{i\sigma'}\quad {\bf M}_{i}^2=m^2 .
\label{letter12}
\end{equation}
Then, the spin-vector of itinerant electrons can be written in the
form
\begin{equation}
{\bf s}_{i}=\frac {1}{2m}{\bf M}_{i}\left(1-h^+_i\,h_i\,-\,
d^+_i\,d_i\right) \label{letter13}
\end{equation}
The contribution of itinerant electrons to the total magnetization
is  $<{\bf s}^z_{i}>$. Accounting for the definition of $m$ (see
Eq.\ref{letter9}), one obtains $<{\bf s}^z_{i}>= <{\bf M}^z_{i}>$.

 The Hamiltonian is quadratic with
respect to the Fermions $d_i, d^+_i$ and $h_i, h^+_i$, and one can
average in the subspace of these Fermions (to integrate them out in
the path integral approach). As a result, we obtain an effective
theory of two spin-vectors ${\bf S}_i$ and ${\bf M}_i$ with
Hamiltonian
\begin{equation}
 h_{eff}= -  J^l\sum\limits_{  \langle  ij  \rangle  } {{\bf S}_i \cdot
{\bf S}_j}-  J^{it}\sum\limits_{  \langle  ij  \rangle  } {{\bf M}_i
\cdot {\bf M}_j}
  - J\sum\limits_i {{\bf
S}_i}\cdot {\bf M}_i \label{letter14}
\end{equation}
The first term is the term which describes the exchange of localized
spins in the Hamiltonian Eq.(\ref{letter1}). The second term is
obtained integrating out the Fermions. It is calculated in the one
loop approximation and in the limit when the frequency and the wave
vector are small. For the effective exchange constant $J^{it}$, at
zero temperature, we obtained
\bea\label{letter14a} & & J^{it} = \\
& & \frac {t}{6m^2}\frac 1N
\sum\limits_{k}\left(\sum\limits_{\nu=1}^3\cos
k_{\nu}\right)\left[1-\theta(-\varepsilon^h_k)+\theta(-\varepsilon^d_k)\right]
\nonumber \\
& - & \frac {2t^2}{3m^2 s J}\frac 1N
\sum\limits_{k}\left(\sum\limits_{\nu=1}^3\sin^2
k_{\nu}\right)\left[1-\theta(-\varepsilon^h_k)-\theta(-\varepsilon^d_k)\right]\nonumber\eea
where $N$ is the number of lattice's sites, $\varepsilon^h_k$ and
$\varepsilon^d_k$  are Fermions' dispersions, \bea\label{letter14b}
\varepsilon^h_k & = & 2t(\cos k_x+\cos k_y+\cos k_z)+s t J/2 +\mu  \\
\varepsilon^d_k & = & -2t(\cos k_x+\cos k_y+\cos k_z)+s t J/2 -\mu,
\nonumber \eea
 and wave vector $k$ runs over the first
Brillouin zone of a cubic lattice. The third term in
Eq.(\ref{letter14}) is obtained from the last term in the
Hamiltonian Eq.(\ref{letter1}) using the representation
Eq.(\ref{letter13}) for the spin of itinerant electrons and
Eq.(\ref{letter9}).

We are going to study the ferromagnetic phase of the two-spin system
Eq.(\ref{letter14}) with $J^l>0,\,\,J^{it}>0$, and $J>0$. To proceed we
use the Holstein-Primakoff representation of the spin vectors ${\bf
S}_j(a^+_j,a_j)$ and ${\bf M}_j(b^+_j,\,b_j)$, where $a^+_j,\,a_j$
and $b^+_j,\,b_j$ are Bose fields. In terms of these fields and
keeping only the quadratic terms, the effective Hamiltonian
Eq.(\ref{letter14}) adopts the form \bea\label{letter17}
 h_{eff}& = & s\,J^l\sum\limits_{  \langle  ij  \rangle
 }(a_i^+a_i+a_j^+a_j-a_j^+a_i-a_i^+a_j)\nonumber\\
 & + & m\,J^{it} \sum\limits_{  \langle  ij  \rangle
 }(b^+_ib_i+b^+_jb_j-b^+_jb_i-b^+_ib_j)\\
 & - & J\sum\limits_i
 (\sqrt{sm}\,[a_i^+b_i+b_i^+a_i]-sb_i^+b_i-ma_i^+a_i)\nonumber
 \eea In momentum space representation,
 the Hamiltonian reads
 \be \label{letter18}
 h_{eff} = \sum\limits_{k}\left (\varepsilon^a_k\,a_k^+a_k\,+\,\varepsilon^b_k\,b_k^+b_k\,-
 \,\gamma\,(a_k^+b_k+b_k^+a_k)\,\right),\ee where the dispersions are given by equalities, \be
\label{letter19}
\varepsilon^a_k\,=\,2sJ^l\,\varepsilon_k\,+\,m\,J,\qquad
\varepsilon^b_k\,=\,2mJ^{it}\,\varepsilon_k\,+\,s\,J, \ee
$\varepsilon_k=3-\cos k_x-\cos k_y-\cos k_z$, and
$\gamma\,=\,J\,\sqrt{s\,m}$.

To diagonalize the Hamiltonian, one introduces Bose fields
$\alpha_k,\,\alpha_k^+,\,\beta_k,\,\beta_k^+$, \be\label{letter21}
a_k\,=\cos\theta_k\,\alpha_k\,+\,\sin\theta_k\,\beta_k,\quad
b_k\,=\,-\sin\theta_k\,\alpha_k\,+\,\cos\theta_k\,\beta_k \ee with
coefficients of transformation, \be\label{letter22}
\cos\theta_k\,=\,\sqrt{\frac 12\,\left (1+\frac
{\varepsilon^a_k-\varepsilon^b_k}{\sqrt{(\varepsilon^a_k-\varepsilon^b_k)^2+4\gamma^2}}\right
)},\ee and $\sin\theta_k=(1-\cos^2\theta_k)^{1/2}$. The transformed
Hamiltonian  \be \label{letter23} h_{eff} = \sum\limits_{k}\left
(E^{\alpha}_k\,\alpha_k^+\alpha_k\,+\,E^{\beta}_k\,\beta_k^+\beta_k\right),
\ee where \be  \label{letter24} E^{\pm}_k\,=\,\frac 12\,\left
[\varepsilon^a_k+\varepsilon^b_k\,\pm \,
\sqrt{(\varepsilon^a_k-\varepsilon^b_k)^2+4\gamma^2}\right] \ee, and
$E^{\alpha}_k=E^{+}_k,\,\,E^{\beta}_k=E^{-}_k$. With positive
exchange constants $J^l>0,\,J^{it}>0$, and $J>0$, the bose fields'
dispersions are positive $\varepsilon^a_k>0,\,\,\varepsilon^b_k>0$
for all wave vectors $k$. As a result, $E^{\alpha}_k>0$ and
$E^{\beta}_k\geq 0$ with $E^{\beta}_0=0$. Near the zero wave vector,
$E^{\beta}_k\approx \rho k^2$ where the spin-stiffness constant is
$\rho=(s^2 J^l\,+\,m^2 J^{it})/(s+m)$. Hence, $\beta_k$ is the
long-range \textbf{(magnon)} excitation in the two-spin effective
theory, while $\alpha_k$ is a gapped excitation with gap
$E^{\alpha}_0\,=\,(s+m)J$.
%\begin{center}
%\begin{figure}[htb]
%\label{letterfig2}
%\centerline{\psfig{file=SWT.eps,width=7cm,height=5cm}}
%\caption{(color online)\,Temperature dependence of the ferromagnetic
%moments: $M$ (blue line)-the magnetization of the system, $M^l$
%(green line)-contribution of the localized electrons, $M^{it}$ (red
%line)-contribution of the itinerant electrons for parameters
%$s=1,\,m=0.2,\,J^l/J=0.5$ and $J^{it}/J=0.1$:\, spin-wave theory}
%\end{figure}
%\end{center}

The dimensionless magnetization per lattice site of the system $M$
is a sum of the  magnetization of the localized electrons
$M^l\,=\,<S^z_i>$ and the magnetization of the itinerant electrons
$M^{it}\,=\,<s^z_i>\,=\,<M^z_i>$,\, $(M\,=\,M^l\,+\,M^{it})$. By
means of the Holstein-Primakoff representation the magnetizations
adopts the form $M^l=s-1/N \sum\limits_{k}<a^+_k a_k>,\,\,
M^{it}=m-1/N \sum\limits_{k}<b^+_k b_k>$. Finally, by means of the
transformation Eq.(\ref{letter21}) one can rewrite $M^l$ and
$M^{it}$ in terms of the Bose functions of the excitations
$\alpha_k$-$n^{\alpha}_k$ and $\beta_k$-$n^{\beta}_k$ \bea
\label{letter26}& & M^l\,=\,s-\frac 1N \sum\limits_{k}\left
[\cos^2\theta_k\,n^{\alpha}_k\,+\,\sin^2\theta_k\,n^{\beta}_k
\right],
\\
& & M^{it}\,=\,m-\frac 1N \sum\limits_{k}\left
[\sin^2\theta_k\,n^{\alpha}_k\,+\,\cos^2\theta_k\,n^{\beta}_k\right].\nonumber
\eea The magnetization of the system is \be \label{letter27}
M\,=\,s\,+\,m-\frac 1N \sum\limits_{k}\left
[n^{\alpha}_k\,+\,n^{\beta}_k\right].\ee

The magnon excitation $\beta_k$ in the effective theory
Eq.(\ref{letter14}) is a complicate mixture of the transversal
fluctuations of the spins of localized and itinerant electrons
Eq.(\ref{letter21}). As a result, the magnons' fluctuations suppress
in a different way the magnetic order of these electrons.
Quantitatively, this depends on the coefficients $\cos\theta_k$ and
$\sin\theta_k$ in Eqs.(\ref{letter26}). If the
spin-Fermion interaction is very strong, $J\gg J^{it}$ and $J\gg
J^l$, one can calculate the coefficients approximately using
approximate expressions for dispersions Eqs. (\ref{letter19}):
$\varepsilon^a_k\approx m J, and \varepsilon^a_k\approx s J$. As a
result, one obtains $\cos^2\theta_k\approx m/(m+s)$. For large $J$,
the gap of the $\alpha$ excitation is very big,
$E^{\alpha}_0\,=\,(s+m)J$, and one can drop this excitation in the
calculations. Then, the approximate expressions for magnetization
satisfy $ M^l/s\,=M^{it}/m$, which means that the strong spin-Fermion 
interaction aligns the magnetic orders of the itinerant and
localized electrons so strong  that they become zero at one and just
the same temperature. The result is different if the spin-Fermion
interaction is relatively small. The magnetization depends on the
dimensionless temperature $T/J$ and dimensionless parameters
$s,\,m,\,J^l/J$ and $J^{it}/J$. We consider a theory with spin of the
localized electrons $s=1$ and calculate the parameters of the
effective theory Eq.(\ref{letter14}) in one Fermion-loop
approximation for density of Fermions $n=0.4$ and microscopic
parameter $J/t=12.4$. The result is $m=0.2$ and $J^{it}/J=0.1$.
Finally, we set $J^l/J=0.5$. For these effective parameters, the
functions $M(T/J)$, $M^l(T/J)$, and $M^{it}(T/J)$ are depicted in
Fig.1 The green line is the magnetization of the localized
electrons, the red line is the magnetization of the itinerant
electrons, and the blue line is the total magnetization. The figure
shows that the magnetic order of itinerant electrons (red line) is
suppressed first, at temperature $T^{*}/J=0.5603$. Once suppressed,
the magnetic order cannot be restored at temperatures above $T^{*}$
because of the increasing effect of magnon fluctuations. Hence, the
magnetization of the itinerant electrons should be zero above
$T^{*}$. As is evident from Fig.1, this is not the result within
customary spin-wave theory.
\begin{center}
\begin{figure}[htb]
\label{letterfig2}
\centerline{\psfig{file=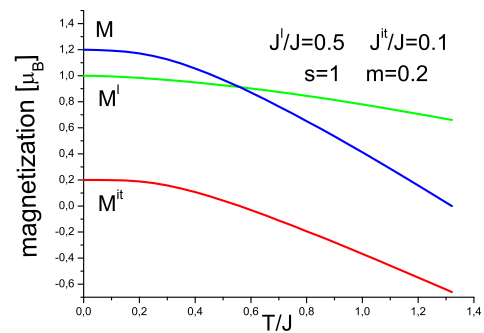,width=6.8cm,height=4.8cm}}
\caption{(color online)\,Temperature dependence of the ferromagnetic
moments: $M$ (blue line)-the magnetization of the system, $M^l$
(green line)-contribution of the localized electrons, $M^{it}$ (red
line)-contribution of the itinerant electrons for parameters
$s=1,\,m=0.2,\,J^l/J=0.5$ and $J^{it}/J=0.1$:\, spin-wave theory}
\end{figure}
\end{center}

To solve the problem, we use the idea on description of paramagnetic
phase of two-dimensional ferromagnets ($T>0$) by means of modified spin-wave
theory \cite{2fmp10,2fmp11}. In the simplest version, the spin-wave
theory is modified by introducing a parameter which enforces the
magnetization of the system to be equal to zero in paramagnetic
phase.

In the present case, we have two-spin system and we introduce two
parameters $\lambda^l$ and $\lambda^{it}$ to enforce the magnetic
moments both of the localized and the itinerant electrons to be
equal to zero in paramagnetic phase. To this end, we add two new terms
to the effective Hamiltonian Eq.(\ref{letter17}), \be
\label{letter28} \hat{h}_{eff}\,=\,h_{eff}\,-\,\sum\limits_i \left
[\lambda^l S^z_i\,+\,\lambda^{it} M_i^z\right]. \ee In momentum
space, the Hamiltonian adopts the form Eq.(\ref{letter18}) with
new dispersions $\hat{\varepsilon}^a_k=\varepsilon^a_k+\lambda^l$
and $\hat{\varepsilon}^b_k=\varepsilon^b_k+\lambda^{it}$, where the
old dispersions are given by equalities (\ref{letter19}). We utilize
the same transformation Eq.(\ref{letter21}) with coefficients
$\cos\hat{\theta}_k$ and $\sin\hat{\theta}_k$ which depend on the
new dispersions in the same way as the old ones depend on the old
dispersions Eq.(\ref{letter22}). In terms of the $\alpha_k$ and
$\beta_k$ bosons, the Hamiltonian $\hat{h}_{eff}$  adopts the form
Eq.(\ref{letter23}) with dispersions $\hat{E}^{\alpha}_k$ and
$\hat{E}^{\beta}_k$, which can be written in the form
Eq.(\ref{letter24}) replacing $\varepsilon^a_k$ and
$\varepsilon^b_k$ with $\hat{\varepsilon}^a_k$ and
$\hat{\varepsilon}^b_k$.

We have to do some assumptions for parameters $\lambda^l$ and
$\lambda^{it}$ to ensure correct definition of the two-boson
theory. For that purpose, it is convenient to represent the
parameters $\lambda^l$ and $\lambda^{it}$ in the form $
\lambda^l\,=\,m J \mu^l\,-\,m J, and \lambda^{it}\,=\,s J
\mu^{it}\,-\,s J$. In terms of the parameters $\mu^l$ and
$\mu^{it}$, the dispersion reads
$\hat{\varepsilon}^a_k=2sJ^l\,\varepsilon_k+mJ\mu^l,\,\,\,
\hat{\varepsilon}^b_k=2mJ^{it}\varepsilon_k+sJ\mu^{it}$ The
conventional spin-wave theory is reproduced when
$\mu^l=\mu^{it}=1$($\lambda^l=\lambda^{it}=0$). We assume $\mu^l$
and $\mu^{it}$ to be positive ($\mu^l>0,\,\mu^{it}>0$). Then,
$\hat{\varepsilon}^a_k>0$, $\hat{\varepsilon}^b_k>0$, and
$\hat{E}^{\alpha}_k>0$ for all values of the wave-vector $k$. To
explore the dispersion $\hat{E}^{\beta}_k=\frac 12 \left
[\hat{\varepsilon}^a_k+\hat{\varepsilon}^b_k-
\sqrt{(\hat{\varepsilon}^a_k-\hat{\varepsilon}^b_k)^2+4\gamma^2}\right]$,
we use the identity $
(\hat{\varepsilon}^a_k-\hat{\varepsilon}^b_k)^2+4\gamma^2=
(\hat{\varepsilon}^a_k+\hat{\varepsilon}^b_k)^2-4(\hat{\varepsilon}^a_k
\hat{\varepsilon}^b_k-\gamma^2)$. It shows that
$\hat{E}^{\beta}_k\geq 0$ if $\hat{\varepsilon}^a_k
\hat{\varepsilon}^b_k-\gamma^2\geq 0$. Since $\hat{\varepsilon}^a_k
\hat{\varepsilon}^b_k\geq
\hat{\varepsilon}^a_0\hat{\varepsilon}^b_0=s m J^2\mu^l \mu^{it}$
for all values of the wave vector $k$, the $\beta_k$ dispersion is
non-negative, $\hat{E}^{\beta}_k\geq0$ if $\mu^l \mu^{it}\geq1$. In
the particular case, $\mu^l \mu^{it}=1$,\,\, $\hat{E}^{\beta}_0=0$,
and, near the zero wave vector, $\hat{E}^{\beta}_k\approx \hat{\rho}
k^2$, with spin-stiffness constant equals  $\hat{\rho}=(s^2 J^l
\mu^{it}+m^2 J^{it} \mu^l)/(s \mu^{it}+m \mu^l)$. Hence, in this
case, $\beta_k$ boson is the long-range excitation (magnon) in the
system. In the case $\mu^l \mu^{it}>1$, both $\alpha_k$ boson and
$\beta_k$ boson are gapped excitations.

We introduced the parameters $\lambda^l$ and $\lambda^{it}$ ($\mu^l,
\mu^{it}$) to enforce the magnetic order of localized and itinerant
electrons to be equal to zero. We find out the parameters $\mu^l$
and $\mu^{it}$ solving the system of two equations $M^l=M^{it}=0$,
where the ordered moments have the same representation as
Eq.(\ref{letter26}) but with coefficients $\cos\hat{\theta}_k,\,\,
\sin\hat{\theta}_k$, and dispersions $\hat{E}^{\alpha}_k,\,\,
\hat{E}^{\beta}_k$ in the expressions for the Bose functions. The
numerical calculations show that
for high enough temperature, $\mu^{it}>1$,\,\, $1>\mu^l>0$, and
$\mu^{it}\cdot\mu^l>1$. Hence, $\alpha_k$ and $\beta_k$ excitations are gapped.
When the temperature decreases, $\mu^{it}$ decreases remaining larger
than one, $\mu^l$ decreases too remaining positive, and  the product
$\mu^l\mu^{it}$ decreases remaining larger than one. At
temperature $T_C/J=2.812$, one obtains $\mu^{it}=5.0427$,
$\mu^{l}=0.1983$, and therefore $\mu^l\mu^{it}=1$. Hence, at $T_C$,
long-range excitation (magnon) emerges in the spectrum which means
that $T_C$ is the Curie temperature.

Below the Curie temperature, the spectrum contains magnon
excitations, thereupon $\mu^l\mu^{it}=1$. It is convenient to
represent the parameters in the following way:
\be\label{letter39}\mu^{it}=\mu, \quad\quad \mu^l=1/\mu.\ee In
ferromagnetic phase, magnon excitations are origin of the suppression
of magnetization. Near the zero temperature, their contribution is
small, and at zero, temperature $M^{it}=m$ and $M^l=s$.  Increasing
the temperature, magnon fluctuations suppress the magnetization. For
the chosen parameters they first suppress the magnetization of the
itinerant electrons at $T^{*}$ ($M^l(T^{*})>0$). Once suppressed,
the magnetic moment of itinerant electrons cannot be restored
increasing the temperature above $T^{*}$. To formulate this
mathematically, we modify the spin-wave theory introducing the
parameter $\mu$ Eq.(\ref{letter39}). Below $T^{*}$,\,\,$\mu=1$,  or
in terms of $\lambda$ parameters $\lambda^l=\lambda^{it}=0$, which
reproduces the customary spin-wave theory. Increasing the
temperature above $T^*$, the magnetic moment of the itinerant
electron should be zero. This is why we impose the condition
$M^{it}(T)=0$ if $T>T^{*}$. For temperatures above $T^*$, the
parameter $\mu$ is a solution of this equation. We utilize the
obtained function $\mu(T)$ to calculate the magnetization of the
localized electrons $M^l$  as a function of the temperature. Above
$T^*$, $M^l$ is equal to the magnetization of the system. The
magnetic moments of the localized and itinerant electrons as well as
the magnetization of the system as a function of the temperature are
depicted in Fig.2 for parameters
$s=1,\,m=0.2,\,J^l/J=0.5, and J^{it}/J=0.1$.
\begin{center}
\begin{figure}[htb]
\label{letterfig5}
\centerline{\psfig{file=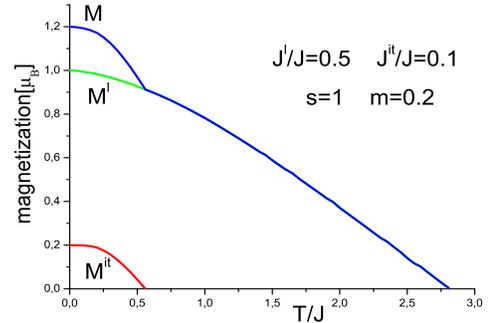,width=6.8cm,height=4.8cm}}
\caption{(color online)\,\,Temperature dependence of the
ferromagnetic moments: $M$ (blue line)-the magnetization of the
system, $M^l$ (green line)-contribution of the localized electrons,
$M^{it}$ (red line)-contribution of the itinerant electrons for
parameters $s=1,\,m=0.2,\,J^l/J=0.5,\,J^{it}/J=0.1$:\,modified
spin-wave theory.}
\end{figure}
\end{center}
The figure shows an anomalous increasing of the magnetization $M$
below $T^{*}$ which is in a very good agrement with the experiment
(see Fig.1 \cite{2fmp6}). The present theory enables us to gain insight
into the nature of the two phases. In the low temperature phase
$(0,T^*)$, the localized and itinerant electrons contribute to the
magnetization of the system, while in the high temperature phase
$(T^*,T_C)$, only localized electrons form ferromagnetic moment. At
first sight, it seems to be counterintuitive because the local
moments build an effective magnetic field, which, due to spin-Fermion
interaction, leads to finite itinerant electron spin polarization.
This is true in the classical limit. In the quantum case, the spin-wave 
fluctuations suppress the magnetic orders of the itinerant and
localized electrons at different temperatures $T^*$ and $T_C$ as a
result of different interactions of the magnon with localized and
itinerant electrons. the spin fermion interaction increases the
alignment of the local moments, and magnetic order of itinerant
electrons is very strong and $T^*$ approaches $T_C$.

It is well known that the onset of magnetism in the itinerant
systems is accompanied with strong anomaly in resistivity
\cite{2fmp12}. This phenomena is experimentally observed at $T^*$ in
the case of $UGe_2$ \cite{2fmp4}. This is another support for the
theoretical interpretation of $T^*$ as a temperature at which the
itinerant electrons form ferromagnetic order.

To conclude, we note that to do more precise fitting with
experimental values of the Curie temperature, one has to account for
the magnon-magnon interaction. However, even the approximate calculations
in the present Brief Report capture the main feature of the two-spin
ferromagnetic systems and the existence of two phases.

The next step of our investigation is to understand the mechanism of
decreasing the phase temperature $T^*$. This will help us to
understand the origin of the superconductivity in these materials.

 This work was financially supported by the Grant-in-Aid for Scientific Research
No19340099 from the JSPS. The author is grateful to Prof. Miyake for
the kind hospitality and useful discussions.


\begin{thebibliography}{99}
\bibitem{2fmp1} S. S. Saxena, P. Agarwal, K. Ahilan, F. M. Grosche, R.K.W. Haselwimmer, M.J.Steiner,
E.Pugh, I.R.Walker, S. R. Julian, P. Monthoux, G. G. Lonzarich, A.
Huxley, I. Sheikin, D. Braithwaite, and J. Flouquet,  Nature
(London) {\bf406}, 587 (2000).
\bibitem{2fmp2} A. Huxley, I. Sheikin, E. Ressouche, N. Kernavanois, D. Braithwaite, R. Calemczuk,
and J. Flouquet, Phys. Rev. B {\bf 63}, 144519 (2001).
\bibitem{2fmp3} N. Tateiwa, T. Kobayashi, K. Hanazono, K. Amaya, Y. Haga, R. Settai,
and Y.Onuki, J. Phys. Condens. Matter {\bf 13}, L17 (2001).
\bibitem{2fmp4} G. Oomi, K. Nishimura, Y. Onuki, and S.W. Yun, Physica {\bf B186-188}, 758 (1993).
\bibitem{2fmp5} N. Tateiwa, K. Hanazono, T. C. Kobayashi, K. Amaya, T. Inoue, K. Kindo,
Y. Koike, N. Metoki, Y. Haga, R. Settai, and Y. Onuki, J. Phys. Soc. Jpn {\bf 70}, 2876 (2001).
\bibitem{2fmp6} C. Pfleiderer and A. D. Huxley, Phys. Rev. Lett., {\bf 89}, 147005 (2002).
\bibitem{2fmp7} G. Motoyama, S. Nakamura, H. Kadoya, T. Nishioka,
and N. K. Sato,Phys. Rev. B {\bf 65}, 020510 (2001).
\bibitem{2fmp8} N. Tateiwa, T. C. Kobayashi, K. Amaya, Y. Haga, R. Setta, and Y. Onuki,Phys. Rev. B {\bf 69}, 180513(R) (2004).
\bibitem{2fmp8a} S. Watanabe and K. Miyake, J. Phys. Society of Japan {\bf 71}, 2489 (2002).
\bibitem{2fmp8b} K. G. Sandeman, G. G. Lonzarich, and A. J. Schofield, Phys. Rev. Lett., {\bf 90}, 167005 (2003).
\bibitem{2fmp8c} S. Q. Shen, International Journal of Physics B {\bf12}, 709 (1998).
\bibitem{2fmp9} D. Schmeltzer, Phys. Rev. B {\bf 43}, 8650 (1991).
\bibitem{2fmp10} M.Takahashi, Prog. Theor. Physics Supplement {\bf 87}, 233 (1986).
\bibitem{2fmp11} M.Takahashi, Phys. Rev. Lett. {\bf 58}, 168 (1987).
\bibitem{2fmp12} P. P. Craig, W. I. Goldburg, T. A. Kitchens, and J.
I. Budnick, Phys. Rev. Lett., {\bf 19}, 1334 (1967).
\end{thebibliography}
\end{document}